\documentclass[12pt]{revtex4}
\usepackage{graphicx}

\begin{document}

\title{Peierls Instability and Electron-Phonon Coupling
in a One-dimensional Sodium Wire}
\author{Prasenjit Sen\footnote{Fax:+91-532-2568036; e-mail:
prasen@hri.res.in}}
\affiliation{Harish-Chandra Research Institute, Chhatnag Road, Jhunsi,
Allahabad 211019, INDIA.}

\begin{abstract}
We have studied Peierls instability in an atomically thin wire of sodium atoms
using first-principles density-functional methods. A Na wire has a stable
uniform linear structure over a range of inter-atomic distances. At smaller
inter-atomic distances it develops a zigzag distortion. At larger
inter-atomic distances, just before
breaking, a Na wire undergoes a very weak Peierls dimerization. This
behavior of a Na wire is understood in terms of its electron-phonon
coupling properties.
\end{abstract}

\maketitle


Fabrication of atomically thin stable wires of gold, and subsequently
those of other $5d$ elements~\cite{ohnishi,yanson} 
have been major breakthroughs
in nanoscience. From the point of view of applications, a striking feature
of these nanowires has been an exact quantization of conductance in units
of the fundamental conductance $2e^2/h$~\cite{ohnishi}. Some recent 
experiments have even measured conductances that are fractions of the 
fundamental unit~\cite{csonka,thijssen}. 
These nanowires hold promise for very
useful applications in nano-electronics and nanotechnology. They
also provide a testing ground for theoretical ideas that have been around for
many years now.

One such important idea is due to Peierls. Peierls argued that a one
dimensional (1D) metallic system, with one valence electron per atom, will
always gain energy by a dimerization of two successive atoms in the
wire~\cite{peierls}.
Such a dimerization creates a potential at a wavevector $2k_F$, and this
opens up a gap in the electronic energy bands. Thus a
1D metallic system is doomed to transform into a Peierls insulator (PI). 
There are many systems, such as MX chains, conjugated polymers,
charge-transfer salts in which Peierls instability plays a major role in
determining their electronic properties~\cite{1dconductor}. 
For systems with bands that are
not exactly half-filled, more complex distortions have been observed,
including simultaneous existence of two different Peierls-like
distortions in Au wires on the Si(553) surface~\cite{Sisurface}. 

First-principles calculations of possible Peierls instabilities in 1D
metallic systems have not always produced conclusive answers.
In an earlier work on Al wires, this author
could not confirm the existence of a Peierls distorted ground state. Later,
Ono and Hirose~\cite{ono} found a spin-Peierls ground state with a six-atom
unit cell for a 1D Al
wire in the regime of large inter-atomic distances, a regime which was not
considered in Ref.~\cite{Sen}. Studies on gold wires have also
been performed. Okamoto and Takayanagi~\cite{okamoto} found that at large
inter-atomic spacings, a dimerized structure is favored even for a
finite Au wire suspended between two electrodes. Conductance of the wire
was found to vanish after dimerization. de Maria and Springborg 
arrived at the same conclusion for an infinite wire of Au
atoms~\cite{demaria}.
Sanchez-Portal {\it et al.}~\cite{sanchez} found only a very small
Peierls gap in their
studies on finite Au wires in the large inter-atomic spacing limit, and 
argued that this instability would play no substantial role in the
properties of Au wires. More recently, Ribeiro and Cohen~\cite{ribeiro} have
studied infinite monatomic wires of Au, Al, Ag, Pd, Rh and Ru. They
concluded that for each of these systems, there is a range of inter-atomic
distances over which a linear wire is stable.

Many of the materials studied so far
are complex in certain respects. Aluminum has three valence
electrons per atom, while for the 5$d$ elements, relativistic effects and spin-orbit
interactions are important. A much simpler, and in fact, an ideal
system to study Peierls instability is a wire of sodium atoms. Sodium is a
simple $sp$ metal with one valence electron per atom. Hence, according to
Peierls's argument, one would expect that a dimerization of its structure
will lower energy. 

In this letter we examine the issue of Peierls instability in a
1D atomically thin infinite wire of Na atoms.
We find that there is a regime of inter-atomic distances where 
a wire of Na atoms does not have a Peierls or 
spin-Peierls instability, and has a stable uniform linear structure with metallic
band structure. At lower interatomic distances, it develops a zigzag distortion
but remains metallic. At higher interatomic distances it undergoes a 
{\em very weak} Peierls transition just before before breaking.
Most significantly, we show that whether a uniform Na wire has a Peierls
instability or not can be rationalized in terms of its electron-phonon coupling
properties. To our knowledge, this is the first independent confirmation 
from first-principles of results obtained in calculations with model Hamiltonians.


First-principles calculations of Na wires are performed within density
functional theory (DFT). The wires are treated within a supercell geometry.
The wire length is taken to be along the $z$ direction. In order to reduce
interactions between a wire and its periodic images in our calculations,
the lengths of the periodic box along $x$ and $y$ are taken to be 20 \AA~ each.
Wavefunctions were expressed in terms of
plane waves with an energy cutoff of 450.0 eV. Potential due to the atomic cores
are represented by Troullier-Martins
pseudopotential. In most of our calculations the exchange-correlation
effects have been treated within local density approximation (LDA) or local
spin-density approximation (LSDA). In a few cases we have treated these
within the Perdew-Burke-Ernzerhof (PBE) 
generalized gradient approximation (GGA). Unless mentioned specifically,
all results refer to LDA/LSDA calculations.
The Brillouin Zone (BZ) integrations were performed within
Monkhorst-Pack (MP) scheme. For ground state electronic structure and geometry
calculations, a $(1\times 1\times 8)$ {\bf k}-point 
grid was used. We checked for
convergence of total energy with respect to the size of the grid. 

Phonon band structure and electron-phonon coupling are calculated within
density functional perturbation theory (DFPT).
For electron phonon calculations, very accurate knowledge of the Fermi surface
is required, and usually a much finer {\bf k}-point grid is necessary. For
this, we used a $(1\times 1\times 24)$ {\bf k}-point grid. Phonon band
structure was obtained by interpolating from energies calculated at 8 {\bf
k}-points. For calculating
electron-phonon coupling strength of bulk Na metal we used a
$(20\times 20\times 20)$ MP {\bf k}-point grid, while the phonon band
structure was interpolated from explicit energy calculations on a $(4\times
4\times 4)$ {\bf k}-point grid. All the numerical calculations are performed
with the ABINIT code.



\begin{figure}[h]
\scalebox{0.4}{ \includegraphics{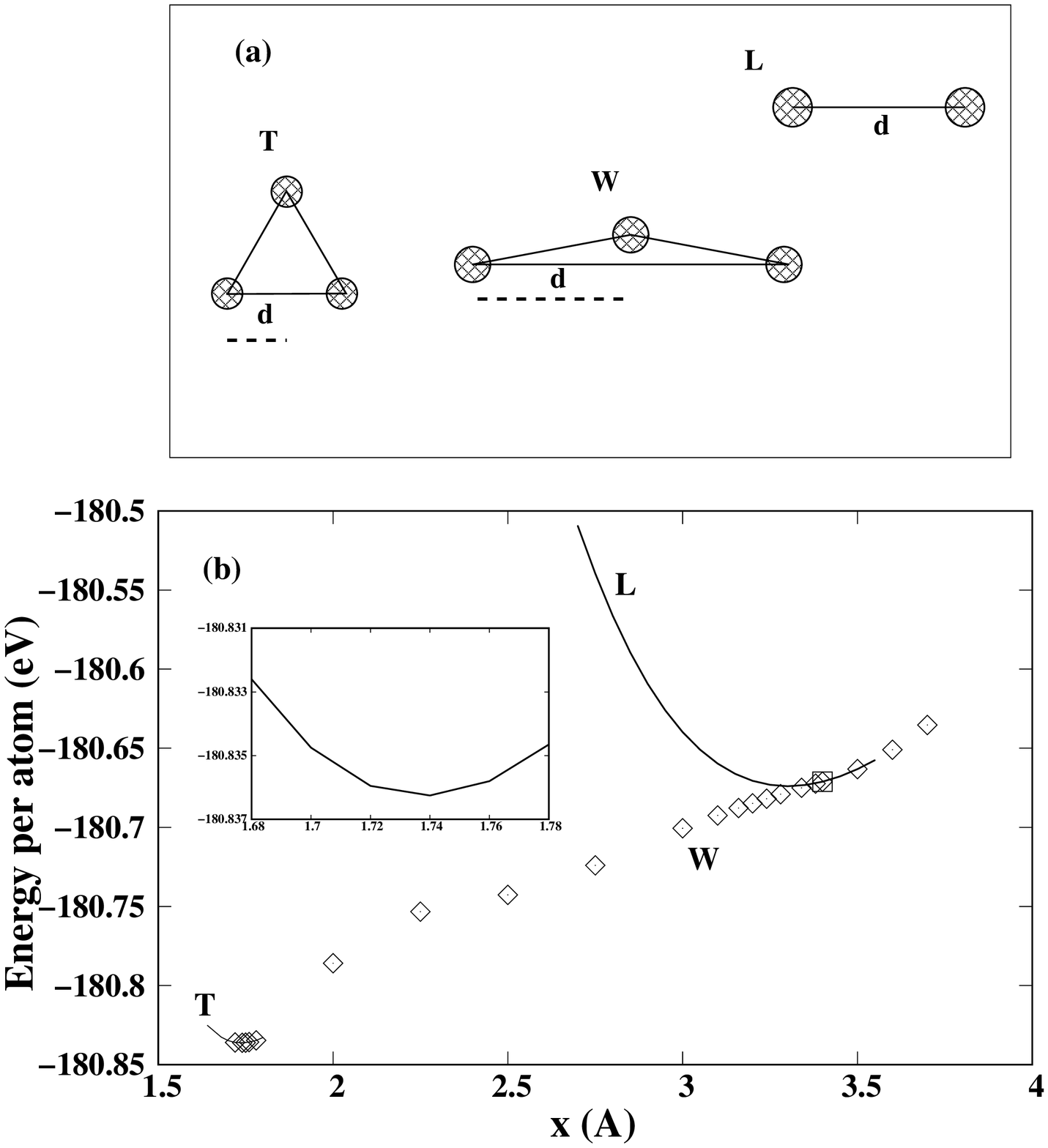} }
\hfill
\caption{(a) Geometries of three different 1D wires of Na atoms
studied in this work.
(b) Variation of total energy per atom as a function of interatomic
distance for {\bf L} (solid line), {\bf W} (diamond),
and {\bf T} (solid line) wires. The inset shows the energy variation of the
{\bf T} wire with $d$.}
\label{fig:structure} 
\end{figure}

Fig.~\ref{fig:structure} plots the total energies of three different 1D
structures of Na atoms 
as a function of inter-atomic distance. Structure {\bf L} is a
uniformly spaced linear wire of Na atoms. {\bf W} and {\bf T} are two
different zigzag structures. The low-energy {\bf T} structure has a bond
angle of $\sim 60^o$, so that three neighboring Na atoms form an
equilateral triangle. Another way of looking at the {\bf T} structure is as
if two parallel uniform linear wires are displaced by half lattice spacing
along the length of the wire. The {\bf W}
structure, on the other hand, has one bond angle that is obtuse at intermediate 
$d$ values, and changes with $d$. In the
{\bf T} structure, each Na atoms has four nearest neighbors (NN), while in the
{\bf W} structure, the number of NN's is two. Presumably, it
is this greater number of NN's that gives the {\bf T} structure a greater
stability. 

The {\bf L} wire has an optimum bond length of $d=3.3$ \AA. However,
the {\bf W} and {\bf T} structures are more favorable
energetically compared to the {\bf L} structure at this, and lower values of $d$. 
It is interesting to compare the behavior of a Na wire with those of other elements.
At lower values of $d$, monatomic wires of all elements studied so far develop zigzag
distortions~\cite{ono,Sen,sanchez,ribeiro,sanchez2}. Of these, only Al and Au wires have a
metastable {\bf W} structure. For all other elements, a {\bf W} wire
transforms to a {\bf T} wire without any energy barrier. In wires
of Ca and K, zigzag distortions set in exactly at the optimum bond length 
of an {\bf L} wire~\cite{sanchez2}. However,
as seen in Fig.~\ref{fig:structure}, a uniform Na wire distorts to a zigzag {\bf W}
wire at $d=3.4$ \AA, even before it reaches its optimum bond length. In this respect a Na wire is
similar to a Cu wire~\cite{sanchez2}. That a Na wire behaves similar to a Cu wire, and is different
from a K wire is rather surprising, and the reason is not immediately clear.

\begin{figure}[h]
\scalebox{0.4}{ \includegraphics{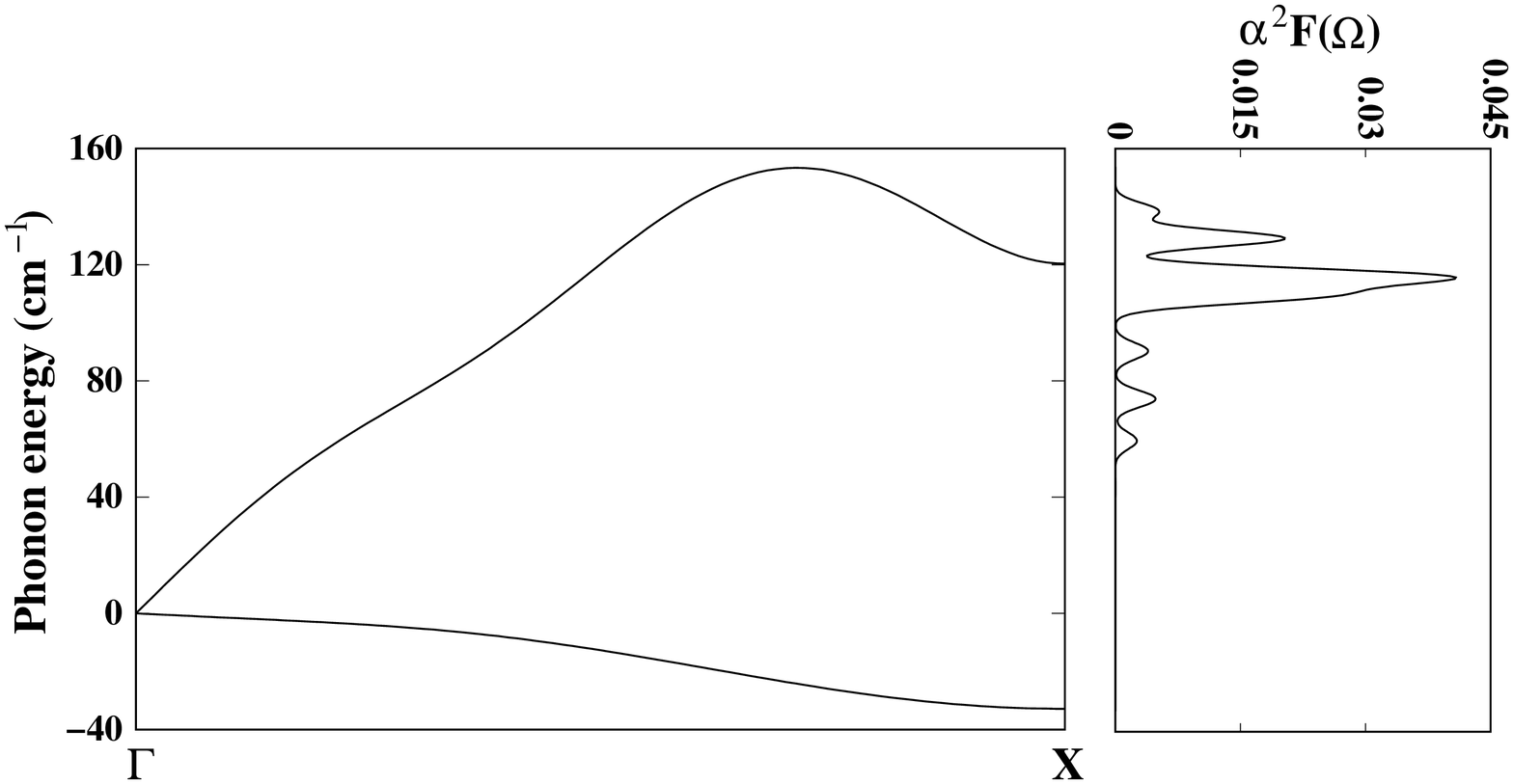} }
\hfill
\caption{Phonon band structure and electron-phonon spectral function
of a {\bf L} wire of Na atoms at $d=3.3$ \AA.}
\label{fig:phonon3.3}
\end{figure}

Why the {\bf L} structure spontaneously distorts to a {\bf W} structure
at intermediate values of $d$ can be understood from its phonon properties.
Any structural instabilities of the
{\bf L} wire will show up as soft modes in its phonon band structure.
Peierls instability is one such
possibility which would be signaled by softening of a longitudinal
mode at the zone boundary.
Fig.~\ref{fig:phonon3.3} shows the phonon band structure of the uniform
{\bf L} wire at its equilibrium structure $d=3.3$ \AA. We
find one longitudinal and two soft transverse modes.
Eigenvectors of the transverse modes indicate that they 
correspond to zigzag distortions of the linear wire.
There are two orthogonal directions 
perpendicular to the wire length in which zigzag distortions 
can take place. This makes the mode doubly degenerate. This indicates that
the {\bf L} wire will spontaneously deform into a zigzag structure.
Importantly, there is no indication of a Peierls instability in the phonon
band structure of the {\bf L} wire.


Though there were no indications of a Peierls instability
in the phonon spectrum of the
uniform {\bf L} wire, we did an explicit search for a PI
phase by dimerizing the uniform wire by hand at $d=3.3$, 3.65, 3.8 and 3.9 \AA. 
This is an independent
check of what is already seen in the phonon spectrum calculated via
DFPT methods. As a test of our calculations, we reproduced the spin-Peierls
ground state of an Al wire at $d=5.6$\AA~ as reported in Ref.~\cite{ono}, with
energy gain over a uniform paramagnetic wire 
in excellent agreement with this work.

\begin{figure}[h]
\scalebox{0.4}{ \includegraphics{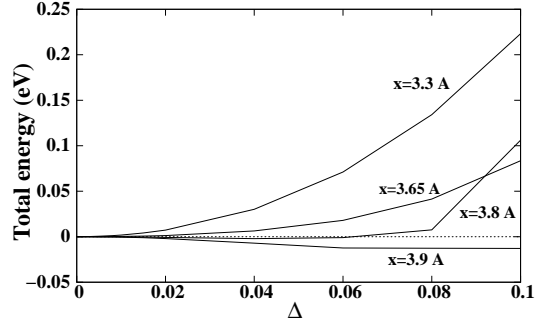} }
\hfill
\caption{Total energy of the {\bf L} wire as a functions of longitudinal
distortion $\Delta$ for different $d$ values. 
In all cases, energy of a uniform wire has been taken
as zero which is denoted by the dotted line.}
\label{fig:distort}
\end{figure}

Fig.~\ref{fig:distort} shows the results of our calculations for
longitudinally dimerized linear wires at the four $d$ values mentioned.
The longitudinal distortion in the uniform wire has 
been quantified by
the parameter $\Delta$, which is equal to the separation of the two
dimerizing atoms normalized by $d$.

At $d=3.3$ \AA, as well as at $d=3.65$ \AA, increasing the magnitude of
longitudinal distortion monotonically increases the total energy of the
wire. At very small distortions, the increase in energy we find is beyond
the limit of accuracy of our DFT calculations. However, the trend is very
clear, and by $\Delta=0.02$, the energy difference with the uniform wire
become significant. This clearly shows that the dimerized structure is not
the ground state of a Na wire at these $d$ vales. However, at $d=3.8$ \AA, 
there is a small decrease in energy due to dimerization over a range
of $\Delta$. With further increase in $\Delta$, the energy eventually increases.
This signifies that the wire has an optimally dimerized stable structure.
The maximum energy gain we find in our LDA calculations is $\sim 1$ meV/atom.
We have repeated this calculation using GGA. The energy gain increases marginally
to $\sim 2$ meV/atom. Although the calculated energy gains are very small, and are barely
at the limit of these calculations, we believe this may be taken as a 
reasonable indication for a PI ground state. This claim is
supported by the phonon band structure of the linear wire at $d=3.8$ \AA~
as discussed later.
This shows that a Na wire will undergo a Peierls transition only at large inter-atomic
separations, and that this instability is 
{\em very weak} in that the energy gained by dimerization is
very small. Moreover, there is a regime of $d$ where
the wire is stable in a uniform {\bf L} structure. Interestingly, at $d=3.9$
\AA, the energy steadily decreases with increasing distortion over the range of
$\Delta$ we have studied. Contrasting this with its behavior at $d=3.8$ \AA,
this is an indication that the wire will break
into isolated dimers at $d=3.9$ \AA.

\begin{figure}[h]
\scalebox{0.5}{ \includegraphics{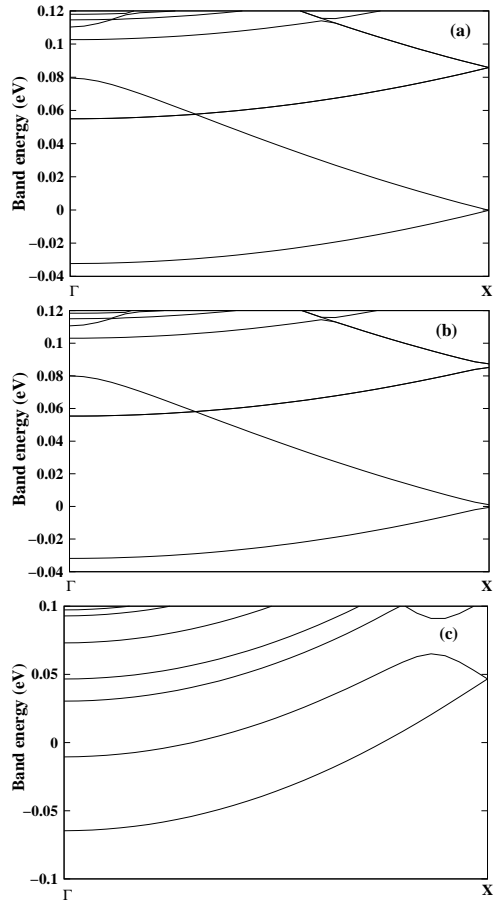} }
\hfill
\caption{Electronic band structures for a uniform {\bf L} (a), 
dimerized {\bf L} at $\Delta=0.01$ (b), and {\bf T} wire (c) of Na atoms. 
The bands of the uniform {\bf L} wire are folded for easy comparison. The
Fermi energy has been set equal to zero in all cases.}
\label{fig:bands}
\end{figure}

Though not the ground state at $d=3.3$ and 3.65 \AA, a dimerized
structure, nonetheless, has two valence electrons per unit cell, and one
would expect it to be an insulator. This is borne out by our
calculations. Fig.~\ref{fig:bands} shows the band structures of a uniform
{\bf L} wire, a dimerized linear wire at $\Delta=0.01$ (both around $d=3.3$
\AA), and the {\bf T} wire in its equilibrium structure at $d=1.74$ \AA. 
To make comparison easier, the bands of the uniform {\bf L} wire have been
folded onto half of the BZ. The uniform wire is metallic with a
half-filled band, as one would expect. The gap opening due to dimerization
is clearly seen in the band structure of the distorted linear wire.
Magnitude of the band gap increases with increase in distortion, $\Delta$.
Interestingly, the {\bf T} structure is also metallic, though it also has 
two valence electrons per unit cell, and naively one would expect it to be
a insulator. It was shown long ago by Batra~\cite{batra} that zigzag
distortions of a 1D metallic system, leading to a doubling of the unit
cell, do not open gaps in their electronic spectrum. It is only the longitudinal
distortions which open gaps. Our calculation is another example of this fact.

\begin{figure}[h]
\scalebox{0.5}{ \includegraphics{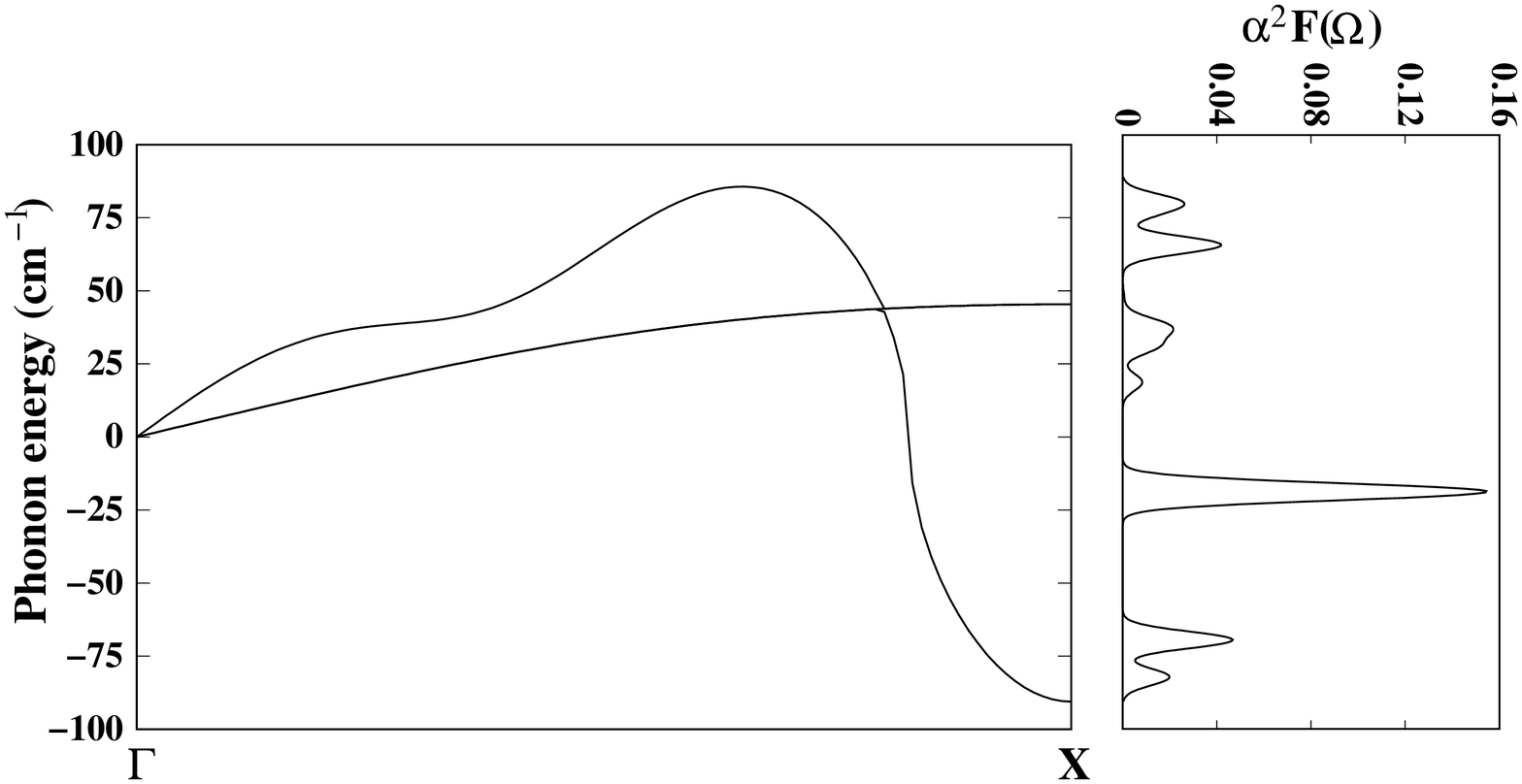} }
\hfill
\caption{Phonon band structure, and electron-phonon spectral function
for a uniform {\bf L} wire at $d=3.8$ \AA.}
\label{fig:phononbnd3.8}
\end{figure}

In order to understand better the structural instabilities of the {\bf L} wire
in the large $d$ regime, we calculated its phonon band structure at $d=3.8$ \AA,
which is shown in Fig.~\ref{fig:phononbnd3.8}. There is one
soft mode in this case near the zone boundary. From the phonon
eigenvectors it is seen to be a longitudinal mode,
which indicates that the uniform {\bf L} wire, in fact, has a Peierls instability
at this $d$. It is also interesting to note that the transverse modes 
giving zigzag distortions of the wire are not soft at this $d$. Consequently, there
is no instability towards zigzag distortions, something we have already seen 
(Fig.~\ref{fig:structure}).

We also searched for anti-ferromagnetic (AFM) and spin-Peierls phases in
a Na wire. Though an $sp$ element like Na is not expected to, and does not
show any magnetic ordering in 3D bulk, it is not completely unexpected for
such elements to show magnetic order at low coordination. In fact, as we
have already mentioned, an Al wire was found to have a spin-Peierls ground
state. We looked for an AFM solution in a uniform {\bf L} wire at $d=3.3$,
3.65, 3.8, and 3.9 \AA, and also looked for spin-Peierls solutions at these
$d$ values. The wires always converged to spin
unpolarized states, thus explicitly confirming that there are no AFM or
spin-Peierls instabilities in a 1D Na wire.


We now try to rationalize why a 1D Na wire has a Peierls instability only at large
values of $d$. For this
we wish to recall a number of calculations that have been
performed on 1D lattice models in the context of Peierls
transition~\cite{bursill,jeckelmann,datta}. 
The three models relevant to our discussions here
are the 1D Holstein model for spin-less Fermions, the same model for
spin-$\frac{1}{2}$ fermions, and the Hubbard-Holstein model. 
A major conclusion from studies on all these models is that a 1D metallic system
undergoes a Peierls transition only if the coupling of the electrons to
phonons is larger than a threshold. While it is
not possible to quantitatively compare values of the parameters in these
lattice models and our first-principles calculations, we think that an explanation
of the absence of
a Peierls transition at smaller values of $d$ may still lie in the electron-phonon
coupling strength in the wire.



Electron-phonon properties of the wires have been calculated within
Migdal theory (MT). A detailed discussion on MT
can be found in Ref.~\cite{SSP}. Whether MT is applicable
to strictly 1D systems is not clear. However, the formalism we have used
treats the simulation system as 3D, with vacuum surrounding the atomically
thin Na wires. In fact, MT has been
applied for calculation of electron-phonon properties in similar
wires~\cite{matthieu} of Al within the same plane-wave DFT formalism. 
Within MT, a measure of the electron-phonon coupling is the
`electron-phonon spectral function' denoted by $\alpha^2 F(\Omega)$.
This is a measure of the effectiveness of phonons of frequency $\Omega$
in scattering electrons from any state to any other state on the Fermi surface.
Figs.~\ref{fig:phonon3.3} and ~\ref{fig:phononbnd3.8} show $\alpha^2 F(\Omega)$
for a uniform {\bf L} wire at $d=3.3$ \AA~ and $d=3.8$ \AA~ respectively, as
a function of phonon energy. It is clear that the electrons primarily couple
to the longitudinal phonon, and that the coupling is more than three
times larger at $d=3.8$ \AA~ compared to that at $d=3.3$ \AA.
Along with electron-phonon coupling strength, $\alpha^2 F(\Omega)$
also depends on the phonon density of states. Averaging over the
phonon density of states, the mass enhancement parameter $\lambda$
is used as a more direct measure of the electron-phonon coupling
strength. $\lambda$ is defined to be 
the first inverse moment of $\alpha^2 F(\Omega)$:
\begin{equation}
  \label{eq:lambda}
  \lambda = 2 \int_0^{\infty} \frac{d \Omega}{\Omega} \alpha^2 F(\Omega).
\end{equation}
We have calculated $\lambda$ for 
the uniform {\bf L} wires at $d=3.3$ \AA~, and $d=3.8$ \AA. For comparison,
we have also calculated its value for bulk Na metal at the optimal lattice
constant calculated within LDA.

\begin{table}
\caption{Calculated and experimental $\lambda$'s for bulk Na metal and
{\bf L} wires}

\begin{tabular}{lcclccr}
\hline
System   &   & & Calc.  &  & & Expt. \\  \hline
Na bulk  &  & &  0.23   & & & 0.22\footnote{Ref.~\cite{elliot}}    \\
{\bf L} wire(d=3.3 \AA)  & & &   0.01  & &  &  -- \\  
{\bf L} wire(d=3.8 \AA)  & & &  0.16  & & &  -- \\  \hline     
\end{tabular}
\label{table:lambda}
\end{table}

Table~\ref{table:lambda} shows our calculated values of $\lambda$ along
with the experimental value for the bulk system. This clearly shows that
the strength of electron-phonon coupling in the 1D {\bf L} 
wire at $d=3.3$ \AA~ is small compared to the bulk system.
More significantly, $\lambda$ increases by an order of magnitude as
$d$ increases from 3.3 \AA~ to 3.8 \AA. We believe, 
it is this increase in electron-phonon
coupling which drives a Na wire to a PI at large values of $d$'s.

In conclusion, we have shown that a monatomic wire of Na atoms has a stable 
uniform linear structure over a range of inter-atomic distances.
Below 3.4 \AA, it distorts to a wide-angled zigzag structure 
which eventually transforms to an equilateral triangular structure on further
compression. On stretching the wire it breaks at $d \sim 3.9$ \AA. Before that,
at $d \sim 3.8$ \AA, it has a PI ground state. However, the Peierls instability
in a Na wire is very weak, as the energy gain on dimerization is very small. We 
rationalize the existence or non-existence of a PI ground state in terms of the
electron-phonon coupling strength in the wire. We also speculate that the
observed stable uniform linear structures of Au, Al, Ag, Pd, Rh, and Ru
over a window of inter-atomic distances~\cite{ribeiro} may have the
same explanation.

The author gratefully acknowledges fruitful discussions with I.\ P.\ Batra
and B.\ C.\ Gupta, and would like thank V.\ Ranjan for his help. All the
computations were carried out on the Tipu cluster at HRI.

\vspace*{0.5in}
\centerline{\bf Figure Captions}

\end{document}